\documentclass[conference]{IEEEtran}

\makeatletter

\def\ps@IEEEtitlepagestyle{%
  \def\@oddfoot{\mycopyrightnotice}%
  \def\@evenfoot{}%
}
\def\mycopyrightnotice{%
  {\footnotesize 979-8-3503-2781-6/23/\$31.00~\copyright~2023 European Union\hfill}
  \gdef\mycopyrightnotice{}
}

\usepackage{blindtext}
\usepackage{eso-pic}
\IEEEoverridecommandlockouts
\usepackage{cite}
\usepackage{amsmath,amssymb,amsfonts}
\usepackage{algorithmic}
\usepackage{graphicx}
\usepackage{textcomp}
\usepackage{xcolor}
\def\BibTeX{{\rm B\kern-.05em{\sc i\kern-.025em b}\kern-.08em
    T\kern-.1667em\lower.7ex\hbox{E}\kern-.125emX}}
    
\usepackage{eso-pic}
\newcommand\AtPageUpperMyright[1]{\AtPageUpperLeft{%
 \put(\LenToUnit{0.17\paperwidth},\LenToUnit{-2cm}){%
     \parbox{0.9\textwidth}{\raggedleft\fontsize{8}{11}\selectfont #1}}%
 }}%
\newcommand{\conf}[1]{%
\AddToShipoutPictureBG*{%
\AtPageUpperMyright{#1}
}
}

\usepackage{tikz}
\usepackage{subcaption}

\newtheorem{definition}{Definition}

\newcommand{\bbr}{\mathbb{R}}

\newcommand{\bbn}{\mathbb{N}}

\newcommand{\indicator}[1]{1_{\{ #1 \}}}
\newcommand*{\bigtimes}{\mathop{\raisebox{-.5ex}{\hbox{\huge{$\times$}}}}}

\begin{document}
\title{\vspace*{1cm} A Comparison of Different Representations of Ordinal Patterns and Their Usability in \\ Data Analysis
\thanks{This work was supported by the German Research Foundation (DFG), grant number SCHN 1231-3/2.}
}

\author{\IEEEauthorblockN{Alexander Schnurr}
\IEEEauthorblockA{\textit{Department Mathematik} \\
\textit{University of Siegen}\\
Siegen, Germany \\
schnurr@mathematik.uni-siegen.de}
\and
\IEEEauthorblockN{Angelika Silbernagel}
\IEEEauthorblockA{\textit{Department Mathematik} \\
\textit{University of Siegen}\\
Siegen, Germany \\
silbernagel@mathematik.uni-siegen.de}}

\maketitle
\conf{\textit{  III. International Conference on Electrical, Computer and Energy Technologies (ICECET 2023) \\ 
16-17 November 2023, Cape Town-South Africa}}
\begin{abstract}
We describe and analyze different approaches to represent ordinal patterns. All of these can be found in the literature. The most important representations (plus sub-classes) are compared in terms of their applicability from different angles. Namely we consider digital implementation, inverse patterns and ties between values. At the end we provide a guideline on which occasions which representation should be used. 
\end{abstract}


\section{Introduction}
Since the seminal paper \cite{bandtpompe2002} by Bandt and Pompe, so called ordinal patterns have been used extensively in contexts of data analysis, dynamical systems as well as time series analysis and mathematical statistics. 
Applications include, but are not limited to 
estimating the Hurst parameter in long range dependent time series \cite{sinnkeller11} calculating entropies of dynamical systems \cite{bandtkellerpompe02, bandtpompe2002, kellersinn10},
analyzing the dependence structure between time series \cite{schnurrdehling17, schnurrfischer22, schnurrfischerties}
and
checking for independence in a series of random variables \cite{weiss22, weissschnurr23}.
Ordinal patterns describe the order of the values in a data set (vector) of length $d$. There are various ways on how to encode ordinal patterns. Sometimes authors pick a certain representation, telling their readers, why it is useful in the context they have in mind. Most of the time, however, one gets the impression, that the representation was chosen randomly or only because `others have used it before'. In this note, we give an overview on the most important representations of ordinal patterns and describe their advantages and disadvantages in different contexts. We also consider variants of the representations. In our comparison we focus on digital implementation, inverse patterns and the handling of ties. 

The paper is organized as follows: In Section II we present the main mathematical concepts. In Sections III to V the representations are compared from different angles. A case study is provided in Section VI and a short conclusion in Section VII rounds out the paper. 

\section{Definitions and mathematical framework}
\label{section: definitions}

For $d \geq 2$, let $S_d$ denote the set of permutations of $\{1, ..., d\}$. Note that $|S_d|=d!$. Furthermore, let $\chi$ be a totally ordered space and $x = (x_1, ..., x_d) \in \chi^d$ with pairwise distinct elements. For now, we focus onto this restriction. Allowing ties, i.e., equalities between some elements, requires more refined representations. This is the subject of Section \ref{section: ties}.

\begin{definition}
    The \emph{ordinal pattern} (of length $d$) of $x$ is defined as the description of the relation of the elements of $x$ in terms of position and rank order.
\end{definition}

\begin{figure}[b]
    \centering
    \begin{tikzpicture}[scale=0.7]
        \draw[gray, thick] (0, 1) -- (1,3);
        \filldraw[black] (0,1) circle (1.5pt);
        \filldraw[black] (0.5,2) circle (1.5pt);
        \filldraw[black] (1,3) circle (1.5pt);
        
        \draw[gray, thick] (2, 1) -- (2.5,3) -- (3, 2);
        \filldraw[black] (2,1) circle (1.5pt);
        \filldraw[black] (2.5,3) circle (1.5pt);
        \filldraw[black] (3,2) circle (1.5pt);

        \draw[gray, thick] (4, 2) -- (4.5,1) -- (5, 3);
        \filldraw[black] (4,2) circle (1.5pt);
        \filldraw[black] (4.5,1) circle (1.5pt);
        \filldraw[black] (5,3) circle (1.5pt);
        
        \draw[gray, thick] (6, 2) -- (6.5,3) -- (7, 1);
        \filldraw[black] (6,2) circle (1.5pt);
        \filldraw[black] (6.5,3) circle (1.5pt);
        \filldraw[black] (7,1) circle (1.5pt);
        
        \draw[gray, thick] (8, 3) -- (8.5,1) -- (9, 2);
        \filldraw[black] (8,3) circle (1.5pt);
        \filldraw[black] (8.5,1) circle (1.5pt);
        \filldraw[black] (9,2) circle (1.5pt);
        
        \draw[gray, thick] (10, 3) -- (11, 1);
        \filldraw[black] (10,3) circle (1.5pt);
        \filldraw[black] (10.5,2) circle (1.5pt);
        \filldraw[black] (11,1) circle (1.5pt);
    \end{tikzpicture}
    \caption{Ordinal patterns of length $d=3$.}
    \label{fig: OP 3}
\end{figure}
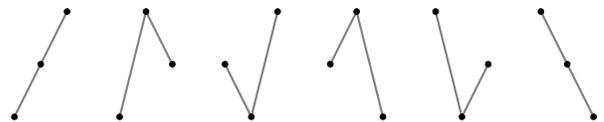

Considering the vector $(5, 3, 7) \in \bbn^3$ as an example, its ordinal pattern is fully specified by: \textit{``Of three elements, the third is the largest, while the second is the least''}. 
Note that this description applies to any vector $(x_1, x_2, x_3)$ satisfying the relation $x_2 < x_1 < x_3$, hence all of them correspond to the same ordinal pattern. 
For $d=2$, there are only 2 ordinal patterns, namely, the upward pattern $x_1 < x_2$ and the downward pattern $x_1 > x_2$. For $d=3$, there are already 6 possible patterns, which are depicted in Fig. \ref{fig: OP 3}.

Now, this specification via text or illustration is quite cumbersome in practice, so the need for other representations arises.

\begin{definition}[Permutation Representation]
    The \emph{permutation representation} of the ordinal pattern of $x$ is defined as the permutation $\pi = (\pi_1, ..., \pi_d) \in S_d$ satisfying
    \begin{equation}
        x_{\pi_1} < ... < x_{\pi_d}.
        \label{eq: permutation representation}
    \end{equation}
\end{definition}

This representation is used, e.g., in \cite{weiss22, weisstestik22, bandtkellerpompe02}.
The tuple defined above consists of the respective indices sorted from the least to the largest value. 
Instead of an increasing order, the permutation representation of ordinal patterns is sometimes defined in a decreasing order, i.e., the pattern $\pi$ has to satisfy $x_{\pi_1} > ... > x_{\pi_d}$ (see, e.g., \cite{schnurrfischer22}). Moreover, even though for our purpose consideration of the length $d$ of an ordinal pattern is more convenient, regarding ordinal time series analysis it is more common to count the number of increments instead of $d$ \cite{betkenetal}, so, starting in zero instead of one, it is prevalent to consider permutation representations $(\pi_0, \pi_1, ..., \pi_d)$ of vectors $(x_0, x_1, ..., x_d)$ with $d \in \bbn$ (see, e.g., \cite{betkenetal, kellersinnemonds, schnurrdehling17, bandtpompe2002}). This already results in four different possibilities to represent an ordinal pattern which are all closely linked to each other. Let us now come to a fundamentally different representation.

\begin{definition}[Rank Representation]
    \label{def: rank representation}
    The \emph{rank representation} $r=(r_1, ..., r_d) \in S_d$ of the ordinal pattern of $x$ is defined by the condition
    \begin{equation}
        r_j < r_k \iff x_j < x_k \text{ for all } j, k \in \{1, ..., d\}.
        \label{eq: rank representation}
    \end{equation}
\end{definition}

This representation is used, e.g., in\cite{bandtshiha07, weiss22, bandt2019, weissschnurr23}. Originally, the permutation representation has been most prevalent in the literature. Now it seems that there is a shift in recent publications as the rank representation seems to become more popular. A reason for this may be the intuitiveness of the concept of ranks in general.

For the rank representation, the entries are given by the ranks of the respective values, where 1 denotes the minimum and $d$ the maximum. Again, an inverted definition in terms of ranks is possible such that 1 denotes the maximum, while $d$ denotes the minimum, though it is not very common. 

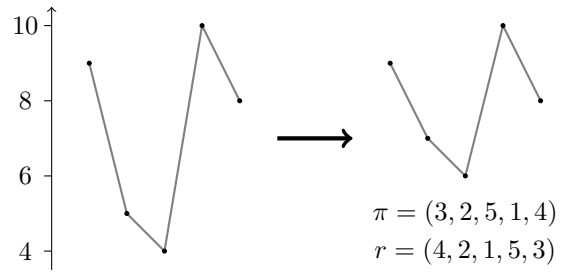
\begin{figure}[b]
    \centering
    \begin{tikzpicture}[scale=0.5]
        \draw[gray, thick] (1,5) -- (2,1) -- (3,0) -- (4,6) --(5,4);
        \filldraw[black] (1,5) circle (1.5pt);
        \filldraw[black] (2,1) circle (1.5pt);
        \filldraw[black] (3,0) circle (1.5pt);
        \filldraw[black] (4,6) circle (1.5pt);
        \filldraw[black] (5,4) circle (1.5pt);

        \draw[->] (0, -0.5) -- (0, 6.5);
        \foreach \x in {4, 6, 8, 10}{
            \draw (0, \x-4) -- (-0.2, \x-4);
            \node at (-0.7, \x-4) {$\x$};
        };

        \draw[->, ultra thick] (6, 3) -- (8, 3);

        \draw[gray, thick] (9,5) -- (10,3) -- (11,2) -- (12,6) --(13,4);
        \filldraw[black] (9,5) circle (1.5pt);
        \filldraw[black] (10,3) circle (1.5pt);
        \filldraw[black] (11,2) circle (1.5pt);
        \filldraw[black] (12,6) circle (1.5pt);
        \filldraw[black] (13,4) circle (1.5pt);

        \node at (11, 1) {$\pi = (3, 2, 5, 1, 4)$};
        \node at (11, 0) {$r = (4, 2, 1, 5, 3)$};
    \end{tikzpicture}
    \caption{Ordinal pattern representations for $x = (9, 5, 4, 10, 8)$.}
    \label{fig: OP representations}
\end{figure}

For better illustration of the ideas, consider Fig. \ref{fig: OP representations}. There the ordinal pattern of the vector $x=(9, 5, 4, 10, 8)$ is depicted together with the respective permutation and rank representations.
Note the dichotomy: While for the rank representation unique ranks are assigned to the indices of $x$, for the permutation representation of ordinal patterns indices are assigned to ranks, so the vector is sorted. Nevertheless, the definitions are equivalent: Any pattern $\pi = (\pi_1, ..., \pi_d)$ can be determined by a distinct permutation function $\sigma : \bbn \to \bbn$ defined by $\sigma(j) = \pi_j$, that is, the condition $x_{\sigma(1)} < ... < x_{\sigma(d)}$ is satisfied. Moreover, its inverse function $\sigma^{-1} : \bbn \to \bbn$ satisfies 
\begin{equation}
    r_j = \sigma^{-1}(j) < \sigma^{-1}(k) = r_k \iff x_j < x_k,
\end{equation}
which already constitutes the rank representation \cite{bergeretal}. 

\begin{definition}[Inversion Representation]
    Define the set
    \begin{equation}
        \mathcal{I}_d = \bigtimes^d_{k=1} 
        \{0, 1, ..., d-k\}
    \end{equation}
    for $d \geq 2$. The \emph{inversion representation} of the ordinal pattern of $x$ is defined as the tuple $i=(i_1, ..., i_d) \in \mathcal{I}_d$ consisting of the (right) inversion counts, that is, 
    \begin{equation}
        i_j = \sum^{d}_{k=j+1} \indicator{x_j > x_k}
        \label{eq: inversion representation}
    \end{equation}
    for $j \in \{1, ..., d\}$.
\end{definition}

This representation is used, e.g., in \cite{bergeretal}.
Note that it holds $i_d = 0$ for the rightmost inversion count, which is why most authors omit it in their definitions of the inversion representation. We leave it as it is, so that all representations using tuples have the same length.

Unlike the other two representations using tuples, the inversion representation has no meaningful pictorial interpretation from which one could immediately read the pattern of up and down within the vector.
Originating in discrete mathematics, this representation is typically used with regard to digital implementation of ordinal patterns as it is very convenient there with regard to numerical encoding of ordinal patterns (see Section \ref{section: digital implementation}), though, instead of the (right) inversion counts, sometimes variations in terms of left inversion counts or non-inversion counts are used as e.g., in\cite{kellersinnemonds, kellersinn05}.

The question now arises whether it is possible to find explicit maps which allow to switch from one representation to another one. For the permutation and rank representations, the question is already settled. Noting that the inversion counts $i_j$, $j \in \{1, ..., d\}$, can be obtained via 
\begin{equation}
    i_j = \sum^{d}_{k=j+1} 1_{\{r_j > r_k\}},
    \label{eq: inversion by ranks}
\end{equation}
where $r_k$ denotes the rank of $x_k$, it is sufficient to show that the permutation representation $\pi$ can be obtained by the inversion representation $(i_1, ..., i_d)$. For this, let $\pi^{(1)}, \pi^{(2)}, ..., \pi^{(d)} = \pi$ be a sequence of permutations of $\{d\}, \{d, d-1\}, ..., \{d, d-1, ..., 1\}$, respectively. Since $\{d\}$ consists of one element, $\pi^{(1)} = (d)$ denotes the trivial permutation. Suppose $\pi^{(l-1)} = (\rho_1, \rho_2, ..., \rho_{l-1})$ is given for an $l \in \{2, ..., d\}$. Then $\pi^{(l)}$ can be obtained from $\pi^{(l-1)}$ by inserting $d+1-l$ into $(\rho_1, ..., \rho_{l-1})$:
\begin{enumerate}
    \item If $i_{d+1-l}=0$, then $d+1-l$ is inserted to the left of $\rho_1$, such that $\pi^{(l)} = (d+1-l, \rho_1, ..., \rho_{l-1})$.
    \item Otherwise it is inserted to the right of $\rho_{i_{d+1-l}}$.
\end{enumerate}
For a similar procedure regarding a definition of the inversion representation in terms of different (non-)inversion counts, see, e.g., \cite{kellersinnemonds}. In fact, this procedure is based on the authors' ideas.

\section{Digital Implementation}
\label{section: digital implementation}

With regard to digital implementation, in order to derive fast algorithms for the determination of ordinal patterns stemming from a time series, it is necessary to keep computational and memory costs (in the sense of storing the obtained ordinal patterns) as low as possible. Therefore, the naive solution of storing ordinal patterns as $d$-dimensional arrays is disadvantageous, since, among other reasons, testing a pair of ordinal patterns for equality would require up to $d$ comparisons, and storing arrays in general results in a way larger memory footprint compared to simple integers \cite{bergeretal}. Hence, an ordinal pattern representation in terms of a single (non-negative) integer is preferable. 

\begin{definition}[Number Representation]
    For $d \geq 2$, let unique non-negative integers $n \in \bbn_0$ be assigned to the ordinal patterns of $x$ according to some bijective map. We call this a \emph{number representation} of ordinal patterns.
\end{definition}

Note that the map is not further specified, i.e., any bijective map can generate a number representation of ordinal patterns, though for digital implementation it is advantageous if the number representation can be directly computed from any other ordinal pattern representation mentioned before instead of implementing some sort of lookup table \cite{bergeretal}. In this context, approaches/solutions using the inversion representation are already available: 
Let $(i_1, ..., i_d) \in \mathcal{I}_d$ be the inversion representation of an ordinal pattern. \cite{kellersinnemonds} proposed a numerical encoding defined by the relation
\begin{equation}
    n_{KSE} = \sum^d_{j=1} i_j \cdot \frac{d!}{(d-j+1)!} \in \{0, 1, ..., d!-1\},
    \label{eq: KSE}
\end{equation}
where `KSE' refers to the authors. Another approach based on the Lehmer code is proposed by \cite{bergeretal}.
There, ordinal patterns are enumerated by
\begin{equation}
    n_{LC} = \sum^d_{j=1} i_j \cdot (d-j)! \in \{0, 1, ..., d!-1\}.
    \label{eq: LC}
\end{equation}
Note that both maps \eqref{eq: KSE} and \eqref{eq: LC} are bijective \cite{kellersinnemonds, bergeretal}.
The second encoding preserves the lexicographic sorting order of the rank representation of ordinal patterns \cite{bergeretal}, so, reading the representations from left to right, a natural enumeration of ordinal patterns in terms of the deviation from an increasing pattern $r = (1, 2, ..., d)$ is provided. Note that with regard to the inversion representation this is equivalent to $(i_1, ..., i_d) \preccurlyeq (i_1^\ast, ... i_d^\ast)$ if and only if
\begin{equation}
    (i_1, ..., i_d) = (i_1^\ast, ..., i_d^\ast) \text{ or } i_1 = i_1^\ast, ..., i_{k-1} = i_{k-1}^\ast, i_k \leq i_k^\ast
\end{equation}
for some $k \in \bbn$ with $k \leq d-1$ \cite{kellersinnemonds}. For a better understanding consider Table \ref{table: numerical encoding}, where both numerical encodings for all possible ordinal patterns of length $d=3$ are listed. Note that even though \cite{kellersinnemonds} claimed their numerical encoding to be lexicographic, here it is not as e.g. Table \ref{table: numerical encoding} shows. Rather it follows a different order. This is due to the different definitions of the vector $x$ under consideration: The vector $x$ considered by \cite{kellersinnemonds} is of the form $x = (x_{-1}, x_{-2}, ..., x_{-d}) = (x^\ast_d, x^\ast_{d-1}, ..., x^\ast_1)$. Therefore, one could argue that the inversion representation the authors use is with respect to non-inversion counts rather than (right) inversion counts. In fact, consideration of variants of inversion counts, as, e.g., left inversion counts or non-inversion counts, does not result in a different type of encoding, but changes the produced enumeration order.
\begin{table}[t]
    \caption{Numerical Encodings for Ordinal Patterns with length $d=3$}
    \centering
    \begin{tabular}{|c|c|c|c|}
        \hline
        Rank Representation&Inversion Representation&$n_{KSE}$&$n_{LC}$ \\
        \hline
        (1, 2, 3) &(0, 0, 0) &0 &0 \\
        (1, 3, 2) &(0, 1, 0) &3 &1 \\
        (2, 1, 3) &(1, 0, 0) &1 &2 \\
        (2, 3, 1) &(1, 1, 0) &4 &3 \\
        (3, 1, 2) &(2, 0, 0) &2 &4 \\
        (3, 2, 1) &(2, 1, 0) &5 &5 \\
        \hline
    \end{tabular}
    \label{table: numerical encoding}
\end{table}

As the encodings presented above only vary with regard to the weights, at first glance one may say that mentioning both encodings is redundant. However, the approach based on the Lehmer code leads to a remarkably simple algorithm for extracting
and storing ordinal patterns in computer memory, which results in a reduction of computational complexity \cite{bergeretal}. 

Algorithms for extracting ordinal patterns from time series data using the encodings presented above have been proposed by \cite{kellersinnemonds}, \cite{unakafovakeller} and \cite{bergeretal}. For a thorough discussion of these algorithms in terms of strengths and weaknesses we especially refer to \cite{bergeretal}.

In summary, the inversion representation is very advantageous with regard to digital implementation of ordinal patterns, since it allows for a direct computation of number representations. What about the permutation and rank representations? 

The rank representation is disadvantageous for digital implementation, since the identification of the rank of an entry needs $d$ comparisons, which sums up to $d^2$ comparisons for the entire representation. In contrast, the identification of the $j$-th inversion number requires $d-j$ comparisons, which results in a total of $\sum^d_{j=1} d-j = (d^2-d)/2$. This can make a major difference regarding the computational time with respect to extracting ordinal patterns from a large data set. Furthermore, (at least to our knowledge) it is not possible to find weights $w_1, ..., w_d \in \bbn$ such that the encoding $\text{en} : S_d \to \{0, 1, ..., d!-1\}$ defined by $\sum^{d}_{j=1} r_j \cdot w_j$ constitutes a bijective map.

Moreover, the permutation representation is unsuitable, too, since even the extraction of the ordinal pattern of a vector makes the additional step of rank identification necessary.

\section{Inverse Ordinal Patterns}

For a vector $x = (x_1, ..., x_d)$ there are two possible ordinal inversions, namely an inversion in time obtained by considering the entries of $x$ in a reversed order, i.e., $(x_d, ..., x_1)$, and an inversion in space given by the reflected vector $(-x_1, ..., -x_d)$. Figuratively speaking, an inversion in time means reflecting on a vertical line, while a reflection on a horizontal line yields an inversion in space. These result in altered ordinal patterns and hence, they have different effects on the proposed representations, which we want to discuss in this section. 
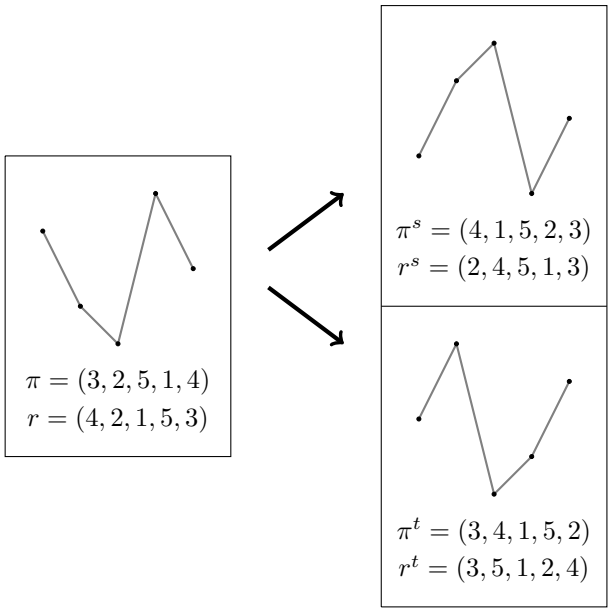
\begin{figure}[b]
    \centering
    \begin{tikzpicture}[scale=0.5]
        \draw[gray, thick] (-1,4) -- (0,2) -- (1,1) -- (2,5) -- (3,3);
        \filldraw[black] (-1,4) circle (1.5pt);
        \filldraw[black] (0,2) circle (1.5pt);
        \filldraw[black] (1,1) circle (1.5pt);
        \filldraw[black] (2,5) circle (1.5pt);
        \filldraw[black] (3,3) circle (1.5pt);

        \node at (1, 0) {$\pi = (3, 2, 5, 1, 4)$};
        \node at (1, -1) {$r = (4, 2, 1, 5, 3)$};

        \draw[->, ultra thick] (5, 3.5) -- (7, 5);
        \draw[->, ultra thick] (5, 2.5) -- (7, 1);

        \draw[gray, thick] (9,6) -- (10,8) -- (11,9) -- (12,5) -- (13,7);
        \filldraw[black] (9,6) circle (1.5pt);
        \filldraw[black] (10,8) circle (1.5pt);
        \filldraw[black] (11,9) circle (1.5pt);
        \filldraw[black] (12,5) circle (1.5pt);
        \filldraw[black] (13,7) circle (1.5pt);

        \node at (11, 4) {$\pi^s = (4, 1, 5, 2, 3)$};
        \node at (11, 3) {$r^s = (2, 4, 5, 1, 3)$};
        
        \draw[gray, thick] (9,-1) -- (10,1) -- (11,-3) -- (12,-2) -- (13,0);
        \filldraw[black] (9,-1) circle (1.5pt);
        \filldraw[black] (10,1) circle (1.5pt);
        \filldraw[black] (11,-3) circle (1.5pt);
        \filldraw[black] (12,-2) circle (1.5pt);
        \filldraw[black] (13,0) circle (1.5pt);

        \node at (11, -4) {$\pi^t = (3, 4, 1, 5, 2)$};
        \node at (11, -5) {$r^t = (3, 5, 1, 2, 4)$};

        \draw (-2, 6) -- (4, 6) -- (4,-2) -- (-2,-2) -- (-2,6);
        \draw (8,-6) -- (8,10) -- (14,10) -- (14,-6) -- (8,-6);
        \draw (8,2) -- (14,2);
    \end{tikzpicture}
    \caption{Ordinal patterns of $x = (9, 5, 4, 10, 8)$ (left) inversed in space (top right) and time (bottom right).}
    \label{fig: OP inversions}
\end{figure}

Inverse patterns (in space) were analyzed in \cite{schnurropd} and \cite{schnurrdehling17}. The appearance of a certain pattern in one data set and the inverse pattern in another data set at the same instants of time can be interpreted as an antimonotone behaviour between these data sets: If at a certain time the value of one of the data sets is high, the value of the other data set is likely to be low. This was first analyzed for financial data like the S\&P 500 and the corresponding volatility index VIX in \cite{schnurropd}. In \cite{sinnkeller11} inverse patterns (in time and space) were used, to make an estimator for the Hurst parameter more efficient by using a Rao-Blackwellization. 

First, let us consider representations of ordinal patterns inversed in space. For such an inversion, the ordinal pattern is reflected on a horizontal line such that the largest value becomes the least, the second largest becomes the second least, and so on. For the permutation representation $\pi = (\pi_1, ..., \pi_d)$ this means that it has to be read from right to left, that is, the permutation representation of the space-inversed ordinal pattern is given by $\pi^s = (\pi_d, ..., \pi_1)$ (or $\pi_j^s = \pi_{d+1-j}$ for all $j \in \{1, ..., d\}$). In contrast, the space-inversed rank representation $r^s=(r^s_1, ..., r^s_d)$ results from an inversion of the type $r^s_j = d+1 - r_j$, $j \in \{1, ..., d\}$. 
For time-inversed ordinal patterns, the entries of $x$ are considered from right to left instead of from left to right, so the time-inversed rank representation is given by the original rank representation read from right to left, i.e., $r^t = (r_d, ..., r_1)$, while the time-inversed permutation representation is obtained via $\pi^t = d+1-\pi_j$ for $j \in \{1, ..., d\}$. Note that the rank representations follow the respective transformations of the vector $x$ under consideration, while the `opposite' transformation (in terms of horizontal and vertical) yields the respective permutation representations. 

Finally, let us consider inversion representations of ordinal patterns inversed in space and time. From \eqref{eq: inversion by ranks}, for the inversion counts $i_j^s$ of the ordinal pattern inversed in space it follows
\begin{align*}
    i_j^s &= \sum^d_{k=j+1} \indicator{r^s_j > r^s_k} 
    = \sum^d_{k=j+1} \indicator{d+1-r_j > d+1-r_k} \\
    &= \sum^d_{k=j+1} \indicator{r_j < r_k} = d-j - i_j
\end{align*}
for $j \in \{1, ..., d\}$, since there are $d-j$ values to compare with $r_j$ and we already know that $i_j$ of these comparisons constitute inversions. Obviously, this can be easily computed from the inversion counts. On the other hand, for ordinal patterns inversed in time it holds
\begin{align*}
    i_j^t &= \sum^d_{k=j+1} \indicator{r_j^t > r_k^t} = \sum^d_{k=j+1} \indicator{r_{d+1-j} > r_{d+1-k}} \\
    &= \sum^{d+1-j-1}_{l=1} \indicator{r_{d+1-j} > r_l}
\end{align*}
for $j \in \{1, ..., d\}$. These constitute the left non-inversion counts, which we cannot deduce from $i_j$ in general.

\section{Ties}
\label{section: ties}

Up to this point, we assumed that ties are not present in $x$. This is an assumption often made in the literature. With regard to time series models, this matches the case that the probability of coincident values equals zero. With regard to data, however, one might still encounter ties. In practice, three approaches are common \cite{schnurrfischerties}:
In the first approach, the respective data points are skipped, i.e., the vectors containing ties are omitted. This is the approach used, e.g., in \cite{bandtkellerpompe02, bandt2019}. In this case one might lose a lot of information, especially considering categorial data sets with a small number of categories. The second approach is randomization, e.g. by adding a small noise to the data in order to avoid ties, which has been done, e.g., by \cite{bandtpompe2002, bandtshiha07}. This has the drawback of possibly underestimating co-movement between data sets or disregarding constant patterns in one data set. 

The last of those approaches is an alteration of the respective ordinal pattern representations using tuples in terms of adding a supplementary condition, that is, then the permutation representation of an ordinal pattern of $x$ is defined as the permutation $\pi = (\pi_1, ..., \pi_d)$ satisfying
\begin{equation}
    x_{\pi_1} \leq ... \leq x_{\pi_d} \text{ and } \pi_{j-1} < \pi_j \text{ if } x_{\pi_{j-1}} = x_{\pi_j}
    \label{eq: permutation representation ties}
\end{equation}
for $j \in \{2, ..., d\}$. This representation is used, e.g., in \cite{betkenetal, weiss22, amigoetal},
and it is equivalent to the adjusted rank representation $r = (r_1, ..., r_d)$ defined by 
\begin{equation}
    r_j < r_k \iff x_j < x_k \text{  or  } (x_j = x_k \text{ and } j < k)
    \label{eq: rank representation ties}
\end{equation}
by the same argument as in the case of distinct values considered in Section \ref{section: definitions}, and which in turn is used, e.g., in \cite{weiss22, weisstestik22}. This means in particular that both representations are used in \cite{weiss22}. Note the additional conditions in comparison to \eqref{eq: permutation representation} and \eqref{eq: rank representation}. 

Since the increasing order is retained in case of ties, the inversion representation does not need to be adjusted further, and equivalence of all three representations using tuples is guaranteed. However, this approach maps the vectors $(1, 1, 1)$ and $(1, 10, 100)$ onto the same pattern $\pi = r = (1, 2, 3)$, thus, they are considered to exhibit the same up and down movement, which clearly is not the case here. Hence, with this approach valuable information is possibly lost. Due to this reason, \cite{schnurrfischerties} propose so called generalized ordinal patterns explicitly allowing for ties by referring to a larger set of possible patterns. The authors propose the following representation, which we will simply refer to as the generalized rank representation:

\begin{definition}[Generalized Rank Representation]
    \label{def: grr}
    Suppose that the values $(y_1, ..., y_m)$, which are already ordered by the condition $y_1 < y_2 < ... < y_m$, are attained in the vector $x = (x_1, ..., x_d)$. There, $m \in \{1, ..., d\}$ is the number of different values. The \emph{generalized rank representation} of the ordinal pattern of $x$ is defined as the vector $\psi = (\psi_1, ..., \psi_d) \in \bbn^d$ satisfying
    \begin{equation}
        \psi_j = k \iff x_j = y_k.
    \end{equation}
\end{definition}

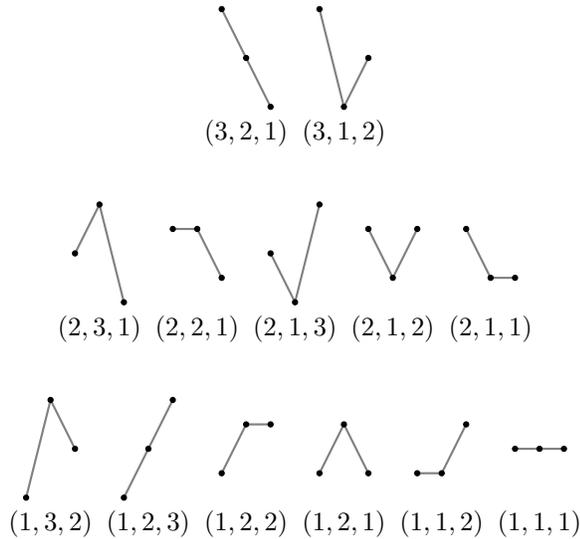
\begin{figure}[t]
    \centering
    \begin{tikzpicture}[scale=0.65]
        \draw[gray, thick] (5,3) -- (5.5,2) -- (6,1);
        \filldraw[black] (5,3) circle (1.5pt);
        \filldraw[black] (5.5,2) circle (1.5pt);
        \filldraw[black] (6,1) circle (1.5pt);
        \node[anchor=south] at (5.5, 0) {$(3, 2, 1)$};

        \draw[gray, thick] (7,3) -- (7.5,1) -- (8,2);
        \filldraw[black] (7,3) circle (1.5pt);
        \filldraw[black] (7.5,1) circle (1.5pt);
        \filldraw[black] (8,2) circle (1.5pt);
        \node[anchor=south] at (7.5, 0) {$(3, 1, 2)$};

        \draw[gray, thick] (2,-2) -- (2.5,-1) -- (3,-3);
        \filldraw[black] (2,-2) circle (1.5pt);
        \filldraw[black] (2.5,-1) circle (1.5pt);
        \filldraw[black] (3,-3) circle (1.5pt);
        \node[anchor=south] at (2.5, -4) {$(2, 3, 1)$};

        \draw[gray, thick] (4,-1.5) -- (4.5,-1.5) -- (5,-2.5);
        \filldraw[black] (4,-1.5) circle (1.5pt);
        \filldraw[black] (4.5,-1.5) circle (1.5pt);
        \filldraw[black] (5,-2.5) circle (1.5pt);
        \node[anchor=south] at (4.5, -4) {$(2, 2, 1)$};

        \draw[gray, thick] (6,-2) -- (6.5,-3) -- (7,-1);
        \filldraw[black] (6,-2) circle (1.5pt);
        \filldraw[black] (6.5,-3) circle (1.5pt);
        \filldraw[black] (7,-1) circle (1.5pt);
        \node[anchor=south] at (6.5, -4) {$(2, 1, 3)$};

        \draw[gray, thick] (8,-1.5) -- (8.5,-2.5) -- (9,-1.5);
        \filldraw[black] (8,-1.5) circle (1.5pt);
        \filldraw[black] (8.5,-2.5) circle (1.5pt);
        \filldraw[black] (9,-1.5) circle (1.5pt);
        \node[anchor=south] at (8.5, -4) {$(2, 1, 2)$};

        \draw[gray, thick] (10,-1.5) -- (10.5,-2.5) -- (11,-2.5);
        \filldraw[black] (10,-1.5) circle (1.5pt);
        \filldraw[black] (10.5,-2.5) circle (1.5pt);
        \filldraw[black] (11,-2.5) circle (1.5pt);
        \node[anchor=south] at (10.5, -4) {$(2, 1, 1)$};

        \draw[gray, thick] (1,-7) -- (1.5,-5) -- (2,-6);
        \filldraw[black] (1,-7) circle (1.5pt);
        \filldraw[black] (1.5,-5) circle (1.5pt);
        \filldraw[black] (2,-6) circle (1.5pt);
        \node[anchor=south] at (1.5, -8) {$(1, 3, 2)$};

        \draw[gray, thick] (3,-7) -- (3.5,-6) -- (4,-5);
        \filldraw[black] (3,-7) circle (1.5pt);
        \filldraw[black] (3.5,-6) circle (1.5pt);
        \filldraw[black] (4,-5) circle (1.5pt);
        \node[anchor=south] at (3.5, -8) {$(1, 2, 3)$};

        \draw[gray, thick] (5,-6.5) -- (5.5,-5.5) -- (6,-5.5);
        \filldraw[black] (5,-6.5) circle (1.5pt);
        \filldraw[black] (5.5,-5.5) circle (1.5pt);
        \filldraw[black] (6,-5.5) circle (1.5pt);
        \node[anchor=south] at (5.5, -8) {$(1, 2, 2)$};

        \draw[gray, thick] (7,-6.5) -- (7.5,-5.5) -- (8,-6.5);
        \filldraw[black] (7,-6.5) circle (1.5pt);
        \filldraw[black] (7.5,-5.5) circle (1.5pt);
        \filldraw[black] (8,-6.5) circle (1.5pt);
        \node[anchor=south] at (7.5, -8) {$(1, 2, 1)$};

        \draw[gray, thick] (9,-6.5) -- (9.5,-6.5) -- (10,-5.5);
        \filldraw[black] (9,-6.5) circle (1.5pt);
        \filldraw[black] (9.5,-6.5) circle (1.5pt);
        \filldraw[black] (10,-5.5) circle (1.5pt);
        \node[anchor=south] at (9.5, -8) {$(1, 1, 2)$};

        \draw[gray, thick] (11,-6) --  (12,-6);
        \filldraw[black] (11,-6) circle (1.5pt);
        \filldraw[black] (11.5,-6) circle (1.5pt);
        \filldraw[black] (12,-6) circle (1.5pt);
        \node[anchor=south] at (11.5, -8) {$(1, 1, 1)$};
    \end{tikzpicture}
    \caption{Generalized rank representations for ordinal patterns of length $d=3$.}
    \label{fig: GRR 3}
\end{figure}

By this definition, the vector $(1, 5, 4, 3)$ yields the generalized rank representation $(1, 4, 3, 2)$, which coincides with the rank representation as defined in Def. \ref{def: rank representation} and is very convenient therefore, while the vector $(1, 1, 4, 3)$ has the generalized rank representation $(1, 1, 3, 2)$ in contrast to $(1, 2, 4, 3)$. Hence, this representation is in fact a generalization of the rank representation with respect to ties. All generalized ordinal patterns of length $d=3$ with their respective generalized rank representations are depicted in Fig. \ref{fig: GRR 3} (in co-lexicographic order).
Note that there are already 13 generalized ordinal patterns of length $d=3$ (and 3 for $d=2$). Denoting the set of all generalized rank representation patterns by $T_d$, the cardinal numbers $|T_d|, d \geq 2$, are given by the Fubini numbers \cite{schnurrfischerties}. 

\cite{schnurrfischerties} applied generalized rank representations to hydrological data sets consisting of five flood classes (plus `absence of flood') in terms of measuring the association within a data set by ordinal pattern dependence. As it is a categorial data set with only six categories, the occurrence of many ties is expected, so the authors compare the proposed generalized ordinal patterns to two classical approaches, namely randomization and the altered definition of sorting from beneath by the first appearance (see \eqref{eq: permutation representation ties} and \eqref{eq: rank representation ties}). Overall, they demonstrate that the classical approaches tend to underestimate the dependence present in the data due to the changes of pattern structure, while their proposed approach of generalized patterns overcomes this. Nevertheless, a new drawback arises: Depending on the length of the ordinal pattern, a lot more patterns need to be considered, which can possibly result, e.g., in a greater computational cost with regard to digital implementation. Therefore, the recommendation is to use this approach when dealing with categorical time series, as it is especially designed for them, while the classical approaches are more advantageous for time series for which the probability of coincident values is very small.

The question now arises to what extent it is possible to find generalizations of the permutation and inversion representation, respectively, which are equivalent to the generalized rank representation as defined in Def. \ref{def: grr}. As a matter of fact, there is a natural way in terms of preimages (in the first case), though notationally disadvantageous: 
The idea of permutation representations is to sort the indices of the entries of $x$ according to their ordinal pattern. However, in case of ties, the indices cannot be sorted in a unique way nor can ties be read directly without further information from the arrangement of the indices in a tuple $\Pi$. This changes when we consider a tuple consisting of sets. Let 
\begin{equation}
    \tilde{\sigma} : \{1, ..., d\} \to \{1, ..., m\}
\end{equation}
be the map that assigns the ranks to the respective indices, where $m$ denotes the previously determined number of different values in $x$. Clearly, $\tilde{\sigma}$ is surjective. Hence, for fixed $l \in \{1, ..., m\}$, the preimage $\tilde{\sigma}^{-1}(l) = \{\psi_1, ..., \psi_{l_n}\}$ is defined, and a generalized permutation representation can be given by 
\begin{equation}
    \Pi = (\tilde{\sigma}^{-1}(1), ..., \tilde{\sigma}^{-1}(m)).
\end{equation}
Consider the following illustrative example: For $x = (4, 4, 6)$, the generalized rank representation is given by $\psi = (1, 1, 2)$, so the first two entries obtain rank 1, while the last entry has rank 2. Then, the generalized permutation representation as proposed above is given by $\Pi = (\{1, 2\}, \{3\})$. This has the advantage that $\{1, 2\} = \{2, 1\}$, so no order is suggested for ties. Nevertheless, the notation is a bit cumbersome compared to sorting the indices themselves as it is done in the classical approach. Obviously, the generalized permutation representation and generalized rank representation are equivalent.

In case of ties, an inversion representation does not make sense as it cannot be defined in a unique way: Considering, e.g., the vectors $(4, 4, 5)$ and $(4, 5, 6)$, under the definition in \eqref{eq: inversion representation} both are mapped to the same inversion representation $(0, 0, 0)$. The naive idea of an adjustment of \eqref{eq: inversion representation} in terms of $\geq$, that is, $i_j = \sum^d_{k=j+1} \indicator{x_j \geq x_k}$, does not solve the problem, since then $(4, 4, 5)$ and $(4, 3, 5)$ are both mapped to $(1, 0, 0)$. The same problem arises when considering other variants of inversion counts. As a consequence, a number representation can be directly computed neither from a generalized inversion representation (as it does not exist) nor the generalized rank representation (due to the same reasons as for the classical rank representation).

\section{Case Study}

In what follows, we consider the S\&P 500 (SPX) and the corresponding volatility index (VIX) as a real-world example in order to illustrate the use of the aforementioned ordinal pattern representations as well as their performance with regard to digital efficiency. We performed our analysis in \texttt{GNU R} on a MacBook Pro (Apple M1).

We have analyzed daily data in the time period 01/02/1990 to 31/01/2023 resulting in $n=8313$ data points, and which is available as open source historical data on finance.yahoo.com. We have restricted ourselves to the `open prices' and, if not mentioned otherwise, we have considered ordinal patterns of length $d=3$. 

\begin{figure}[t]
    \centering
    \begin{subfigure}{0.45\textwidth}
        \centering
        \includegraphics[scale=0.35]{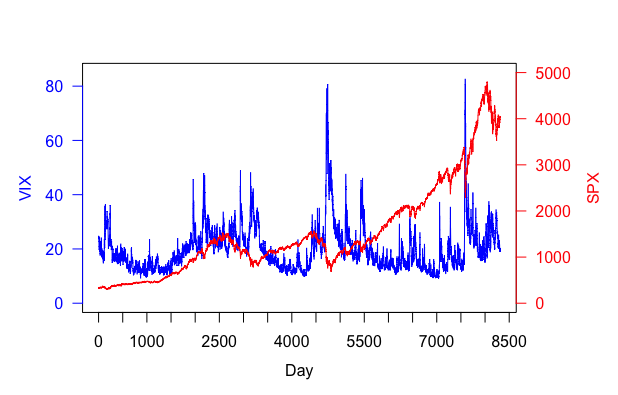}
    \end{subfigure}
    \begin{subfigure}{0.45\textwidth}
        \centering
        \includegraphics[scale=0.35]{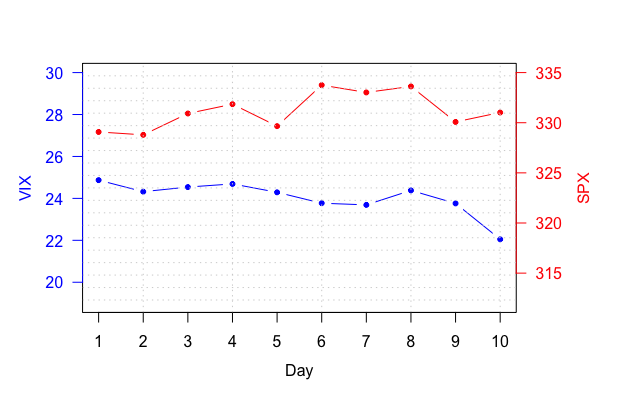}
    \end{subfigure}
    \caption{`Open prices' of VIX and SPX, respectively, in the time period 01/02/1990 to 31/01/2023 corresponding to $n=8313$ data points (top). First 10 data points corresponding to the time period 01/02/1990 to 14/02/1990 (bottom).}
    \label{fig: VIXvsSPX}
\end{figure}

The first 10 data points are illustrated at the bottom of Fig. \ref{fig: VIXvsSPX} from which the respective ordinal patterns can be mapped directly to the rank or permutation representation. If one is interested in the probability of certain patterns, e.g., the upward movement $\pi = r = (1, 2, 3)$ as it might be connected to economic growth (in the case of SPX), determining the relative frequencies for estimation is crucial. Here, we make use of the number representation $n_{LC}$ (see \eqref{eq: LC}), which utilizes indirectly the inversion representation $i = (i_1, .., i_d)$, and use the `Plain Algorithm' as proposed by \cite{bergeretal}. In summary, this yields the relative frequencies stated in Table \ref{table: case study 1}. Using the \texttt{R}-package \texttt{tictoc}, we have obtained the results in far less than one second, respectively. Note that the data exhibits a negligible number of ties, hence, here we have used the altered definition of ordinal patterns which retains the increasing order in case of ties (see Eq. \eqref{eq: permutation representation ties}, \eqref{eq: rank representation ties}), so the inversion representation does not need to be adjusted further.

\begin{table}[b]
    \caption{Relative frequencies of ordinal patterns $n_{LC}$ of length $d=3$ with regard to VIX and SPX rounded to the third digit.}
    \centering
    \begin{tabular}{|c|c|c|c|c|c|c|}
        \hline
        $n_{LC}$& 0 & 1 & 2 & 3 & 4 & 5 \\
        \hline
        VIX & 0.210 &0.135 &0.137 &0.134 &0.131 &0.253 \\
        SPX &0.295 &0.130 &0.127 &0.118 &0.121 &0.209 \\
        \hline
    \end{tabular}
    \label{table: case study 1}
\end{table}
\begin{table}[b]
    \caption{Ordinal pattern dependence (OPD) for different lengths $d$ with regard to VIX and SPX rounded to the third digit.}
    \centering
    \begin{tabular}{|c|c|c|c|c|c|c|}
        \hline
        $d$& 2 & 3 & 4 & 5 & 6 & 7 \\
        \hline
        OPD & -0.366 &-0.251 &-0.149 &-0.075 &-0.037 &-0.018 \\
        \hline
    \end{tabular}
    \label{table: case study 2}
\end{table}

In addition, we have investigated the computational cost with regard to ordinal pattern dependence, which had been proposed together with the \texttt{R}-package of the same name in \cite{schnurropd}. The function \texttt{patterndependence}, which can be found therein, uses an approach for assigning number representations to the respective patterns based on reflected patterns, that is, adding the numbers of reflected patterns has to yield $d! - 1$. Since this is not based on the inversion representation, an additional randomization is required in case of ties. Therefore, slightly different results can be obtained in each run. We have summarized our results for different pattern lengths $d$ in Table \ref{table: case study 2}. All in all, the time needed for the computation of ordinal pattern dependence with $d=3$ is about one third of the time needed with regard to the pattern probabilities computed before, even though these probabilities must also be determined for ordinal pattern dependence. Moreover, the computation of ordinal pattern dependence with $d=7$ takes still far less than one second, but twice the time with regard to the aforementioned computations. This is the case for various reasons: Firstly, parts of the computation in \texttt{patterndependence} are outsourced to \textsc{C} shortening the computation time. Secondly, by using reflected patterns loops can be avoided in the algorithm, which might contribute significantly to the low computational costs at least in case of \texttt{R}. To our knowledge, a comparison of the two algorithms with regard to computational efficiency is not available yet. However, this is beyond the scope of this paper.

\section{Conclusion}

In probability theory and statistics, most of the time the distribution of ordinal patterns is of interest. In order to determine or estimate probabilities of certain ordinal patterns, the pattern representation itself is not relevant, since the choice of the representation usually does not influence the obtained results directly. However, it is still advantageous to opt for a representation for
which the ordinal pattern can be read directly, namely the permutation or rank representation. Here, we prefer the second one, since the concept of ranks is more intuitive and the ups and downs of the ordinal pattern can be read directly.

Another advantage of the rank representation is, that a data vector like $x=(1,3,2,4)$ is mapped on $(1,3,2,4)$ which is very natural. For $\chi=\bbr$ in the absence of ties, every vector $x$ having the pattern $r$ in the rank representation can be derived by applying a strictly monotone transformation $f:\bbr\to\bbr$ on $r$, that is $(x_1,...,x_d)=(f(r_1),...,f(r_d))$. 

When considering inverse patterns, permutation and rank representation are both reasonable, since they behave in a similar, but `opposite' way regarding the transformations needed for the respective inversions (see above). On the other hand, the inversion representation, even though counterintuitive, is advantageous with respect to ordinal patterns inversed in space due to its closed form in terms of the original inversion counts, but with regard to inversions in time, it is inconvenient, because then one has to compute non-inversion counts which cannot be obtained from the (right) inversion counts directly.
However, the inversion representation is very practical when it comes to digital implementation, since it leads to remarkably simple algorithms for extracting and storing ordinal patterns in computer memory, that keep computational and memory costs very low. 

In the context of ties being present in the data, the use of the three classical approaches using tuples is recommended for data sets for which many ties are not expected, e.g., data stemming from a real-valued time series. Then, even for the third approach, where the definitions of the permutation and rank representation are altered, respectively, the previously discussed advantages and disadvantages remain valid. However, if, e.g., a categorial data set with a small number of categories is considered, many ties are to be expected. Hence, the classical approaches would lead to a distortion of the underlying distribution and/or loss of valuable information. Therefore, in this case the generalized rank representation is very beneficial, even though the number of possible patterns increases even more rapid for the length $d$ than the set of permutations considered in the case of vectors containing pairwise distinct elements. The equivalent generalized permutation representation, which we proposed, is notationally more cumbersome, hence, we still recommend the use of the generalized rank representation.


\bibliographystyle{IEEEtran}
\bibliography{IEEEabrv,references}
\vspace{12pt}

\end{document}